\documentclass[pra,twocolumn]{revtex4} 
\usepackage{amssymb}
\usepackage{amsmath}
\usepackage{graphicx}
\usepackage[colorlinks]{hyperref} 
\usepackage{mathrsfs} 

\newcommand{\scrH}{\mathscr{H}}

\newcommand{\ket}[1]{\left | #1 \right \rangle}
\newcommand{\bra}[1]{\left \langle #1 \right |}
\newcommand{\kb}[1]{| #1 \rangle \langle #1 |}

\newcommand{\beq}{\begin{equation}}
\newcommand{\eeq}{\end{equation}}
\newcommand{\beqa}{\begin{eqnarray}}
\newcommand{\eeqa}{\end{eqnarray}}

\begin{document}
\title{Entropic time endowed in quantum correlations}
\author{Akimasa  Miyake}
\affiliation{%
\mbox{Perimeter Institute for Theoretical Physics, 31 Caroline St. N.,
Waterloo ON, N2L 2Y5, Canada} }
\date{November 11, 2011} 

\begin{abstract}

A possible mechanism of time is formulated by developing an idea of time replaced by quantum correlations, 
with the aid of modern quantum information theory.
We invent a microscopic model, where correlations of a closed system are steadily
read out as internal, quantum clocks that define time via their relative phases.
The model could realize emergent time evolutions which exhibit unitarity of quantum theory, 
while its underlying process is driven entropically.
The key quantity turns out to be the amount of accessible information about the clocks recording past events.
By postulating the so-called data-processing inequality (or strong subadditivity of entropy) as a fundamental, 
physical limitation about how information decays,
we propose that conditional entropy about this past information should be constrained to be a positive constant.
The proposal implies a holographic property of this conditional entropy in an analogous manner with the area law of 
entanglement entropy.

\end{abstract}
\pacs{}

\maketitle
\section{Introduction}

``Time is what the constant flow of {\it bits} of sand in an hourglass measures,'' one says.
Although time has been the most fundamental notion to us from ancient days, it could not have been
treated satisfactorily in the modern physics. Time is not an ordinary observable (dynamical variable) in 
quantum theory, nor thus quantized. The time evolution generated by the Einstein equation is 
a mere diffeomorphism of the spacetime, and in this regard it can be said timeless in relativistic theory.

A timeless approach to quantum theory \cite{page83,wootters84} has showed conceptually how an ``evolution'' of the system 
can be observed in a {\it stationary} total state $\ket{\Psi}$.
Imagine the universe consists of two parts:
a hypothetical system $A$ which refers to the coordinate time and a regular system $B$ (which includes the observer). 
They are assumed to be constrained to be stationary under a non-interacting Hamiltonian
$\scrH = h^{A} \otimes {\bf 1}^{B} - {\bf 1}^{A} \otimes h^{B} $; namely, the Wheeler-DeWitt like equation
$\scrH \ket{\Psi}^{AB} = 0$ is satisfied.
We define a clock time when the clock $A$ points its time as $t$ by $\ket{t}^{A} = e^{i h^{A} t} \ket{o}^{A}$, 
with a fiducial zero-time state $\ket{o}^{A}$. (Hereafter, 
the time $t$ should be distinguished from the $\pm 1$ eigenstates, $\ket{0^\mu}$ and $\ket{1^{\mu}}$, of 
the Pauli operator $\sigma^{\mu}$, which appear with the superscript for an axis $\mu$.)
Then, the state $\langle t | \Psi \rangle$ of the system at the time $t$ obeys the standard unitary time evolution 
(with $i = \sqrt{-1}$):
\beq
i \frac{\partial}{\partial t} \langle t | \Psi \rangle = \bra{t} h^{A} \otimes {\bf 1} \ket{\Psi} =
\bra{t} \scrH + {\bf 1}\otimes h^{B} \ket{\Psi} = h^{B} \langle t | \Psi\rangle .
\label{Sch}
\eeq
In other words, formally the quantum correlation between the clock and the system is described as
\beq
\ket{\Psi} = \int dt  \, e^{i h^{A} t} \ket{o}^{A} \otimes  e^{- i h^{B} t} \langle o | \Psi \rangle^{B} . 
\label{AB}
\eeq
This can be interpreted as time emerges from the comparison (i.e. a relative phase) between
two ``precessing'' spins.

However, this is not totally satisfactory in several manners.
First of all, the clock is not treated {\it operationally} in the framework of quantum theory.
Its reference to the time $t$ is not supposed to be monitored quantum-mechanically, nor is 
treated to create information (or entropy) associate with it.
The time flow is rather considered to be absolute (unaffected from any disturbance),
by ticking steadily, globally and deterministically.
Obviously, this is a remote cause to a long-standing puzzle about the nature of macroscopically-observed time, 
such as the second law of thermodynamics. 

Here we develop the above idea of time replaced by quantum correlations \cite{wootters84},
according to recent progress of quantum information and computation theory. 
We invent an operational model, and scrutinize the changes of (both quantum and classical) entropies
in the elapse of time which is intrinsically ruled by the correlations. 
Our guiding rule is one of the most fundamental features in information theory: 
the so-called data-processing inequality \cite{schumacher96} (or mathematically the strong 
subadditivity of entropy).
It states conceptually that no correlation between two parts can be strengthened if they are acted separately. 
Our second key idea is to link this fundamental limitation on entropy changes with the common feature of time 
in quantum theory.
Namely, searched is the condition that the unitary evolution of Eq.~(\ref{Sch}) is realizable {\it steadily}, 
despite the fact that seemingly the global entropy is kept generating.
This brings us an unexpected constraint on certain conditional entropy, as a generalized counterpart 
of the Wheeler-DeWitt constraint.
Physically, it supports an ``area law'' of accessible information about the past events, in that 
it suffices to signal part of the past history in compared with the naive ``volume law'' that all the information 
should be accessible. Needless to say, our area law is quite reminiscent of the conjectured area law of 
entanglement entropy, which bases the holographic picture of our universe.

Finally, we note that our work might be complementarily helpful to understanding why quantum correlation
should be ``limited'' among other conceivable probabilistic theories.
In particular, it may have sounded, out of the blue to physicists, that its Tsirelson bound on nonlocality could be
related to communication complexity \cite{dam05, brassard06} (or, recently to a generalized date-processing 
inequality itself \cite{dahlsten11} among other significant works).
Our work will demonstrate, rather inside quantum theory, innate relationship between the nature of 
time ruled by correlations and the ability of communication (and computation).

\section{Operational, microscopic model}

\begin{figure}[t]
\begin{minipage}{0.49\columnwidth}
\begin{center}
\includegraphics[width=0.95\columnwidth]{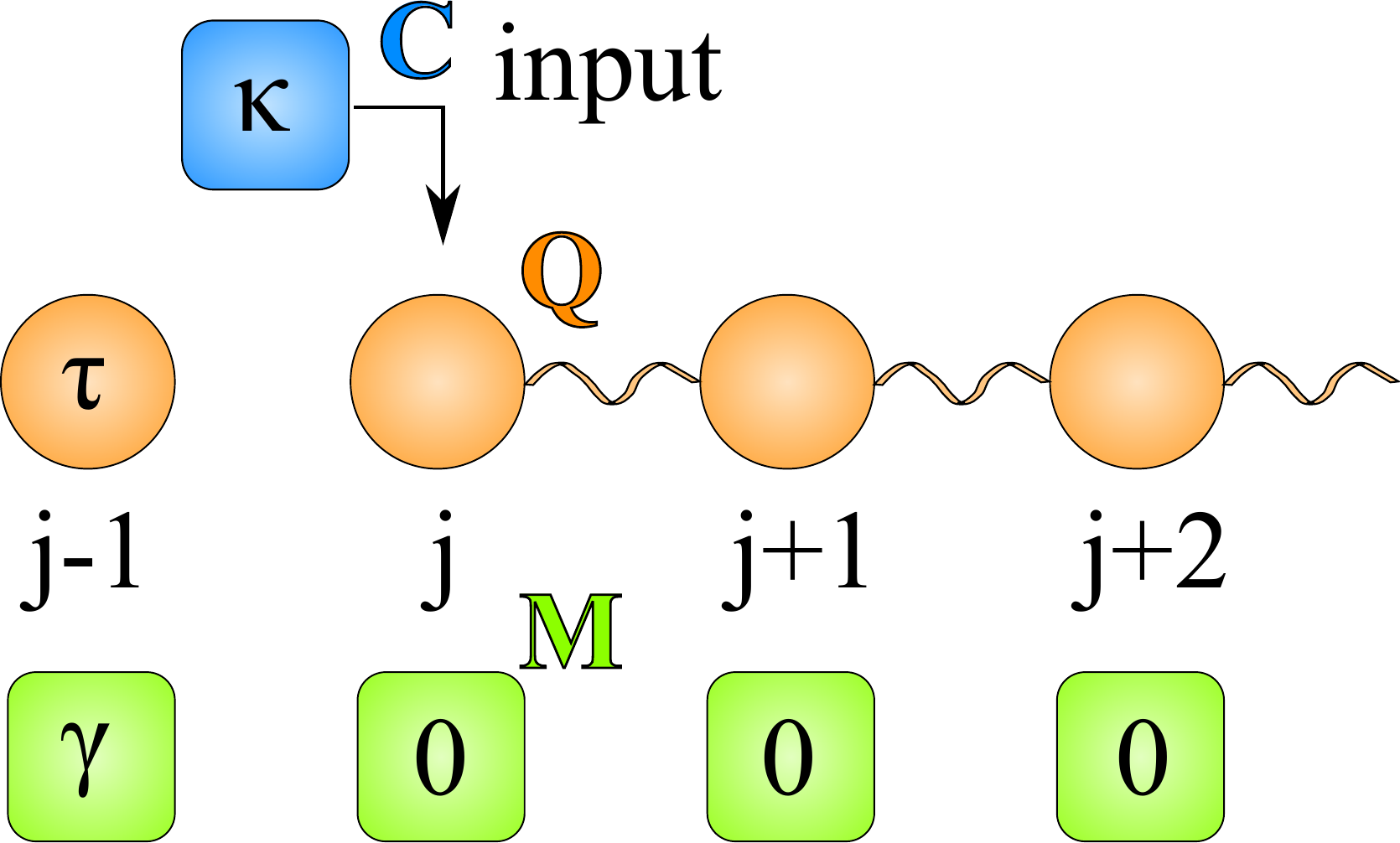}  
\vspace{1cm}
\end{center}
\end{minipage}
\begin{minipage}{0.49\columnwidth}
\begin{center}
\vspace{0.42cm} 
\includegraphics[width=0.95\columnwidth]{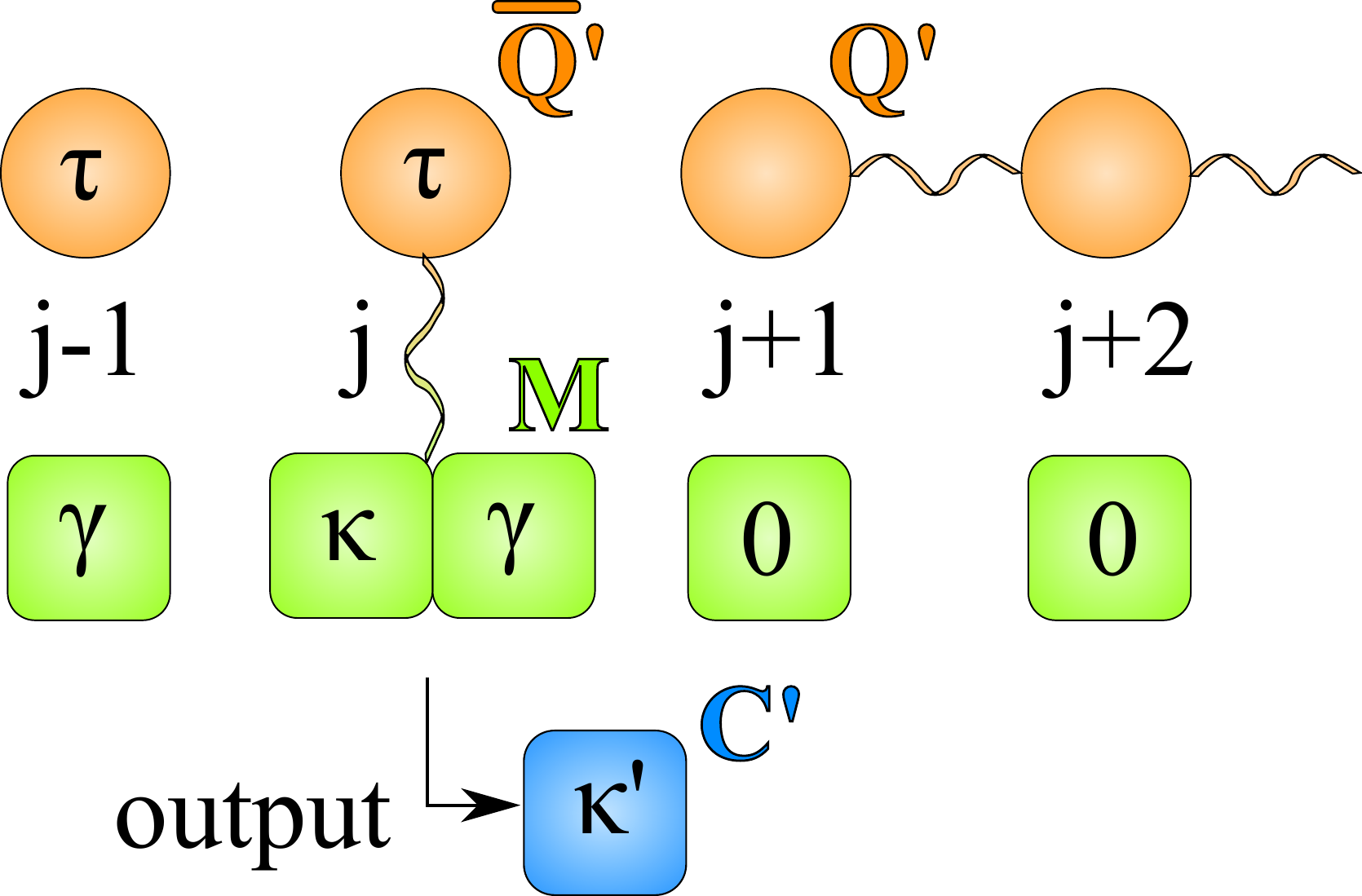}  
\end{center}
\end{minipage}
\caption{A quantum-classical hybrid system that materializes time through correlations among internal clock states.
Every site, which consists of a quantum system and a classical memory, reacts upon receiving an input classical information
by outputting another classical information. A change of the balance of entropies through this elementary process could give
rise to a {\em stationary} information flow to describe time ordering.
These sites are numbered by $j \in \mathbb{N}$ for convenience, but they are not necessarily arrayed in the
one-dimensional manner as seen later. 
(Left) The beginning of the turn when the $j$-th site gets active, after the sites until $j-1$ have been processed.
Based on the input information $\kappa$ of $C$, the $j$-th qudit, which is part of the quantum system $Q$ correlated 
over unprocessed sites, is going to be measured. Its ``yes, no'' outcome $\gamma$ of the measurement is going to be stored 
in the $j$-th classical memory $M$, which is initially uncorrelated with $C$ and $Q$.
(Right) The end of the $j$-th turn. The measured qudit, now denoted as $\bar{Q'}$, points a clock time $\tau$ depending on
the measurement outcome $\gamma$. Part of classical informations $\kappa$ and $\gamma$ is going to send out as
the output $C'$ toward the unprocessed sites, whose quantum part is denoted as $Q'$. 
}
\label{fig:CMQ}
\end{figure}

We are based on a standpoint that laws of nature are about how quantum and classical information can be processed.
Since our idea is to refine the timeless mechanism in an operational and microscopic manner, 
we will naturally consider a quantum-classical {\it hybrid} system.
The entropy of every quantum degree of freedom materializes through an elementary ``yes, no'' phenomenon, 
as advocated e.g., in \cite{wheeler82}. 
It is formulated as consisting of its measurement, outputting only a yes or no
answer (or generally one of discrete alternatives), upon the receipt of input information.
As seen in the Figure~\ref{fig:CMQ}, we model the quantum part $Q$ to be made of an array of $N$ qudits, whose 
local Hilbert space is $d^Q$-dimensional.
It is straightforward to introduce the following two kinds of {\it classical} degrees of freedom.
One is a classical memory $M$ to store the classical outcome of each measurement.
Every degree of freedom in $Q$ is associated with a counterpart in $M$ with the same local size ($d^M := d^Q$).
The other is a classical register $C$ to represent the signaling information to the current state of $Q$,
and is available as the input to the following measurement of $Q$.
The size of $C$, denoted as $d^C$, will be a key parameter.

The current status of the system is described by the density operators $\zeta^Q_{\kappa}$ of $Q$ together with 
classical information $\kappa$ (of at most $\log_2 d^C$ bits) in $C$,
\beq
\rho^{CQ} = \sum_\kappa p_{\kappa} \kb{\kappa}^C \otimes \zeta^{Q}_{\kappa} .
\label{CQ}
\eeq
On the other hand, the reduced state only on $Q$ is
described by $\rho^{Q} = {\rm tr}_{C} \rho^{CQ} = \sum_{\kappa} p_{\kappa} \zeta^{Q}_{\kappa}$.
Importantly, if $C$ is imagined to include all information of $M$, the current state of $Q$ should be able to figure out
its full past history, implying it is described globally as a pure state. 
However, this assumption requires that the size $d^C$ of $C$ scales according to the length of the past,
and then the resulting mechanism is not translationally invariant (i.e., not stationary in the time direction).
What if $C$ can include only the part of information in $M$? 
The past history stored in $M$ lies there in the system, but is considered to be {\it inaccessible} from 
the current standpoint of $Q$. 
That makes $CQ$ to be effectively a mixed state.

The elementary ``measuring'' process is formulated using a local reversible transformation 
$\Upsilon^{CMQ}$ among $CQ$ and the ancillary memory $M$.
$M$ is initially prepared in a fiducial state $\ket{0^z}$ and thus uncorrelated to $CQ$.
Although $\Upsilon^{CMQ}$ will be illustrated explicitly in the Appendix~\ref{app:upsilon} for the simplest case, it induces
some effective quantum action $A(\gamma)$ to update the internal state of $Q$,
depending on the measurement outcome $\gamma$ to be stored in $M$.
It also exchanges locally classical information between $C$ and $M$ to produce an output $C'$. 
Note that the qudit once processed by $\Upsilon^{CMQ}$, which is denoted by $\bar{Q'}$, is
detached from the renewed status $C'Q'$ in a similar manner that the previous $M$ is not accessible to $C'Q'$ anymore.
\begin{align}
\rho^{C' Q'} 
&= {\rm tr}_{M\bar{Q'}} \left[ \Upsilon^{CMQ} (\rho^{CQ} \otimes \kb{0^z}^M) \Upsilon^{\dag CMQ}  \right] 
\nonumber\\
&= \sum_{\kappa' \subseteq \{\kappa, \gamma \}} p'_{\kappa'} \kb{\kappa'}^{C'} \otimes  A(\gamma) \zeta^{Q'}_{\kappa}
A^{\dag}(\gamma) ,
\label{C'Q'}
\end{align}
where $\kappa'$ is part of classical information made of $\kappa$ and $\gamma$.
Thus the current status of the system is described holographically, lying in the interface between the processed region 
of $\bar{Q}$'s and the unprocessed region of $Q$.

\section{Principle of least conditional entropy}

Now we will formulate a global variational principle using the entropies of the bipartite state $C Q$,
and characterize a timeline.
We define the conditional entropy of $C Q$ of Eq.~(\ref{CQ}) by
\begin{align}
S(C | Q) &:=  S(\rho^{CQ}) - S(\rho^{Q})  \nonumber\\ 
	     &=  H(p_{\kappa}) - \left[S(\sum_\kappa p_\kappa \zeta_\kappa) - \sum_{\kappa} p_\kappa S(\zeta_\kappa)\right] ,
\label{condS}	     
\end{align}
where $H(p_{\kappa}) = - \sum_{\kappa} p_\kappa \log p_\kappa$ is the Shannon entropy of the probability
distribution $\{ p_{\kappa}\}$ and
$S(\rho) = - {\rm tr} (\rho \log \rho)$ is the von Neumann entropy of the density operator $\rho$.
Note that $S(C | Q)$ is non-negative in our $CQ$ hybrid system (although the quantity defined in the first line 
of Eq.~(\ref{condS}) can be negative for a general bipartite state).
$- S(C | Q)$ has been sometimes called coherent information related to the channel capacity.
See e.g., \cite{cerf97,horodecki07} more about the properties of the conditional entropy.

The strong subadditivity of the von Neumann entropy says that 
$S(\rho^{12}) - S(\rho^{2}) \geq S(\rho^{123}) - S(\rho^{23})$ holds for any tripartite quantum system $123$ 
(e.g., \cite{ruskai02} for the review, and \cite{short10} to see how unique it is for quantum theory).
By considering that $1= C'$, $2= Q'$, and $3= \bar{Q'}M$ and using that $M$ is initially uncorrelated 
to either of $CQ$, we get a quantum data-processing inequality in our setting:
\beq
S(C | Q) \leq S(C' | Q') .
\label{data-process}
\eeq
It is equivalent to say that at every elementary step, $\Delta S(\rho^{Q}) \leq \Delta S(\rho^{CQ})$ must be satisfied 
regarding the change of entropy $\Delta S$, which is defined as the entropy after the step minus the entropy before.
The data-processing inequality is considered to describe one of the most fundamental features of information.
It can be also said that this is another way to view the entropic uncertainty relation \cite{coles11b,berta10}. 
In our context, given the current state $Q$, missing information about the past events, to be delivered in $C$, cannot 
decrease (see \cite{coles11a} more about properties of this missing information).
Obviously, the best way to preserve the information is to satisfy 
the extremal condition of Eq.~(\ref{data-process}), which will turn out to be the counterpart of the unitary condition
for a quantum-only system.
However, the significance of its extremal value itself may have been overlooked by often looking at a single step.
As pointed out in the previous section, the naive solution to include all information of $M$ in $C$ results in 
its extremal value to be dependent of the whole system size.

Here we propose that the extremal condition should be satisfied not only in a single step but also
{\it every step} in the independent manner of the system size $N$.
Since the behavior is qualitatively similar as far as the extremal value is a constant,
we can pay attention to its least value to be shown as $ \log d^C / d^Q$.
That is how we formulate a global variational requirement,
\beq
S(C|Q) = S(C'|Q') = \log \frac{d^C}{d^Q}  \quad \mbox{for any step} ,
\label{principle}
\eeq
and call it here {\em the principle of least conditional entropy}.
One may probably wonder, at a first glimpse, if the least value could be set to be zero (namely $d^C = d^Q$), 
because of an apparent resemblance of Eq.~(\ref{condS}) to the celebrated bound for the Holevo quantity.
However, the strict positivity in Eq.~(\ref{principle}) seems to be an inevitable feature of time in noncommutative
quantum world, in reflecting the computational universality of quantum theory, discussed later.

\section{Timeline from quantum correlations}

Now we like to specify the quantum correlations of $Q$ in our operational, microscopic model 
as a solution of the proposed principle.
Although the structure of tripartite states which saturates the strong subadditivity inequality has
been analyzed in \cite{hayden04}, we need to introduce some heuristic assumptions to deal with
the sequence of such inequalities as in Eq.~(\ref{principle}).
Inspired by the correlation of Eq.~(\ref{AB}) in the Introduction, 
we assume the primitive bipartite correlation of two qubits (i.e., $d^Q =2$ for simplicity hereafter) is given 
in terms of the stationary subspace of a non-interacting Hamiltonian 
$\scrH =\tfrac{1}{2}( \sigma^z \otimes {\bf 1} - {\bf 1} \otimes \sigma^w )$. 
The first axis can be set to the $z$ axis without loss of generality. 
The second $w$ axis is specified by the relation to the first, $\sigma^w = V \sigma^{z} V^{\dag}$,
where $V$ is generally an element of $SU(2)$ parametrized as 
$ e^{i \theta_4}R^z (\theta_3) R^y(\theta_2) R^z (\theta_1)$ with three Euler angles, using a rotation matrix 
$R^\mu (\theta) = e^{i\tfrac{\theta}{2}} \kb{0^\mu} + e^{-i \tfrac{\theta}{2}} \kb{1^\mu} $.
Here we can set $\theta_1$ to be zero and choose later $\theta_4$ freely, because both are irrelevant 
in the definition of $\sigma^w$. 
The 2-dimensional projector to the stationary subspace is conveniently written as 
\beq
\Xi_t = \oint dt R^{z} (t) \otimes R^{w} (-t) = \kb{0^z 0^w} + \kb{1^z 1^w} ,
\label{Xi}
\eeq
where the integral is taken over $[0, 2 \pi]$ and normalized with $\tfrac{1}{2 \pi}$.
Note that here $t$ is formally an integral variable, originated from geometry of the underlying Hilbert space.
However, this integral representation is insightful to remind us of two {\it correlated} precessing spins seen 
through Eq.~(\ref{AB}).

We define the ``timeline state,'' which handles the history of the time evolution of a single qubit.
Although we hardly use the following explicit form, it is formally written as the simple convolution of our
primitive correlations:
\begin{widetext}
\begin{align}
\Xi_{t_{N-1}} \cdots \Xi_{t_1} (\ket{o}_{1}  \ket{o}_{2} \cdots  \ket{o}_{N})
=   \oint dt_{N-1} \cdots \oint dt_2 \oint dt_1 [ R^z (t_1) \ket{o}_1  \otimes   
& R^z (t_2) R^w (-t_1) \ket{o}_2 \otimes \cdots \otimes R^w (-t_{N-1}) \ket{o}_{N} ] .
\label{timeline}
\end{align}
\end{widetext}
Another motivation for this definition is that, as elaborated in the Appendix~\ref{app:dirac}, the timeline state appears to 
correspond to (the discretized version of) a $1+1$D Dirac fermion, which describes how the spin degree
of freedom should be intertwined with space-time degrees of freedom in a consistent manner with
both quantum theory and relativistic theory.
Conditions for the choice of a fiducial state $\ket{o}$ is derived later.
Depending on $\ket{o}$, we assume the total state is suitably normalized if necessary.
Also notice that every pair of $\Xi_{t_j}$ and $\Xi_{t_{j+1}}$ $(j=1, \ldots , N-2)$ does not commute each other unless $w = z$, 
so that the notation implies that $\Xi_{t_{j+1}}$ acts after $\Xi_{t_j}$.

An elementary process is a projection of the current degree of freedom of $Q$ (the one with the smallest
numbering among unmeasured qubits) onto a certain pointer state $\kb{\tau}$.
Here, based on the principle of least conditional entropy in Eq.~(\ref{principle}), we like to derive
(i) conditions for $\ket{o}$ and $\ket{\tau}$, and (ii) conditions for the $w$ axis.
Using the spin-coherent-state representation, we can parametrize 
$\ket{o} = \cos \lambda_o e^{i\eta_o /2} \ket{0^w} + \sin \lambda_o e^{-i\eta_o /2} \ket{1^w}$ and
$\ket{\tau} = \cos \lambda_{\tau} e^{i\eta_{\tau} /2} \ket{0^z} + \sin \lambda_{\tau} e^{-i\eta_{\tau} /2} \ket{1^z}$.

(i) In analogous manner to the conventional formulation of quantum data-processing inequality 
\cite{schumacher96,nielsen98}, the equality of Eq.~(\ref{principle}) intends that $A$ is unitary
up to a normalization for every outcome $\gamma \,(=0,1)$, corresponding to each clock time $\tau(\gamma)$.
Namely, at the $j$-th step, we get a transfer matrix $A := {\mathstrut}_{j} \!\!\bra{\tau} \Xi_{t_{j}} \ket{o}_{j+1} $,
which should satisfy $A^{\dag} A \propto {\bf 1}$.
Given $\ket{o}$, the unitarity condition is satisfied if $\sin \lambda_{\tau} = \pm \cos \lambda_{o}$.
Furthermore, in order that these two ``outcomes'' constitute a valid measurement (mathematically
a positive operator valued measure: $\sum_{\gamma} \kb{\tau (\gamma)} = {\bf 1}$), they have to obey 
$\sin \lambda_{\tau} = \pm \cos \lambda_{\tau}$.
Note that the azimuth angles $\eta_{o}$ and $\eta_{\tau}$ are not constrained at all. Only the {\it relative}
difference $\tau := \eta_{\tau} - \eta_{o}$ matters in $A$, so that we simply set $\eta_{o} = 0$ and $\eta_{\tau} = \tau$.
That is how we can fix both $\lambda_{o}$ and $\lambda_{\tau}$ to be $\tfrac{\pi}{4}$ without loss of generality, 
by absorbing the possible minus sign as an additional $\pi$ in the azimuth angle. 
In short, for $\ket{o} = \ket{0^w} + \ket{1^w}$ and an arbitrarily fixed $\tau$, 
there are two ``yes or no'' outcomes $ \ket{\tau (\gamma)}  = e^{i \tau/2} \ket{0^z} + 
e^{-i \tau /2}\ket{1^z}$, where $\tau (\gamma)$ is either $\tau$ itself or not $\tau \,(= \tau + \pi)$, each of which 
is labeled by the classical information $\gamma = 0$ or $1$ respectively.

Accordingly, the transfer operator defined above is
\beq
A(\gamma)  \! = \! \oint dt_{j} R^{w} (-t_{j}) \ket{o}_{j+1}   {\mathstrut}_{j} \!\!\bra{\tau (\gamma)} R^{z}(t_j ) 
=  V R^{z}(-\tau (\gamma)),
\label{transfer}
\eeq
where at the second equality we use a key identity, 
\beq
\oint dt |-t \rangle \langle \tau - t| = R^{z} (-\tau) ,
\eeq 
for the canonical clock-time state $\ket{t}:= e^{i t/2} \ket{0^z} + e^{- i t/2} \ket{1^z} $.
We can readily confirm that the normalizations (and thus the probabilities) for the outcomes $\tau$ and $\tau + \pi$
are always equal by construction.
Although it is mathematically straightforward, the identity is indeed based on quantum nature of the clock:
$\langle t | t' \rangle = \cos \tfrac{t-t'}{2}$, in that different times are not orthogonal in allowing 
a chance of a quantum leap.
It would be intriguing to mention that a classical-clock counterpart $\langle t | t' \rangle = \delta (t -t')$,
defined on the continuous degree on ${\mathbb R}$, satisfies formally the same identity.

(ii) Although the unitary transfer operator in Eq.~(\ref{transfer}) allows to satisfy the equality of 
the data-processing inequality in Eq.~(\ref{principle}), it does not reach the suggested
least value $\log \tfrac{d^C}{d^Q}$ in general.
This value accounts for communication cost of the outcomes $\gamma$, often hidden in the conventional formulation with
the classical clock. Indeed, it will turn out that the condition for the $w$ axis (which represents non-commutativity), 
or the counterpart of the ``mass'' term $m (\sim \theta_2)$ by the Dirac-fermionic analogy in the Appendix~\ref{app:dirac}, is 
relevant to it.
It is convenient to simplify further the operator $V$ by using the freedom to choose the $y$ axis.
Actually, $\theta_3$ can be set $\pi$ 
by redefining  $\sigma^{y} := - \cos 2\theta_3 \sigma^{y} - \sin 2\theta_3 \sigma^{x}$ while
maintaining the value of $\theta_2$.
Together with the freedom of $\theta_4$, this leads to the property that $V$ can be chosen hereafter to be Hermitian,
$V = \sigma^{z} e^{i m \sigma^y} = V^{\dag}$.

Let's consider first the homogeneous case ($\tau_j = 0 \, \forall j$).
At the $j$-th stage of Eq.~(\ref{C'Q'}), we would get 
$\widetilde{A} := \prod_{j} A(\gamma_{j}) = \cdots (\sigma^{w})^{\gamma_{4}}(\sigma^{z})^{\gamma_{3}}
(\sigma^{w})^{\gamma_{2}} (\sigma^{z})^{\gamma_{1}}$ 
by an alternating product of $\sigma^w$ and $\sigma^z$ in each branch.
If $m$ is a general angle (or, not a rational multiple of $\pi$), we need all the past classical information
about the outcomes has to be sent to maintain the distinguishability of all the branches and to satisfy 
the extreme of Eq.~(\ref{principle}).
That amounts to that the transferring information satisfies the volume law: $d^C = 2^N$.
Indeed, this abundant communication is hidden in the conventional formalism.

On the other hand, if $m$ is assumed to be suitably ``quantized,'' by being a rational multiple of $\pi$, there is such a chance
that the set of $\widetilde{A}$'s remains effectively a finite number of branches.
Then full transfer of the past may not be necessary, and indeed it is possible to send only a {\it constant} amount 
regardless of the size of the past.
We will discuss in the next Section that this can be interpreted as manifestation of an area laws of the conditional entropy in 1D. 
The necessary and sufficient condition is that $\widetilde{A}$ constitutes a complete depolarization map, in other words
a unitary 1-design, within finite steps.
The simplest solution is given when $m$ is $\tfrac{\pi}{4}$ or its odd multiple, namely $\sigma^w = \pm \sigma^x$ in 
satisfying ${\rm tr} (\sigma^w \sigma^z) = 0$ at the order $2$ (cf. \cite{ambainis00, bouda07}).
At every two steps the equally-weighted branches constitute the Pauli group (disregarding the global phase) 
by ${\bf 1}, \sigma^{z} ,\sigma^{x}, \sigma^{x}\sigma^{z}$ for the set 
$(\gamma_{j+1}, \gamma_{j}) = (0,0), (0,1), (1,0), (1,1)$, respectively.
Then it is sufficient to send only 2 bits ($d^C = 4$), that leads to the least value in Eq.~(\ref{principle}) of
$\log \tfrac{d^C}{d^Q} = 1 $ for the qubit case.

\section{Spacetime structure by a partially ordered geometry}

\begin{figure}[t]
\begin{minipage}{0.44\columnwidth}
\begin{center}
\includegraphics[width=\columnwidth]{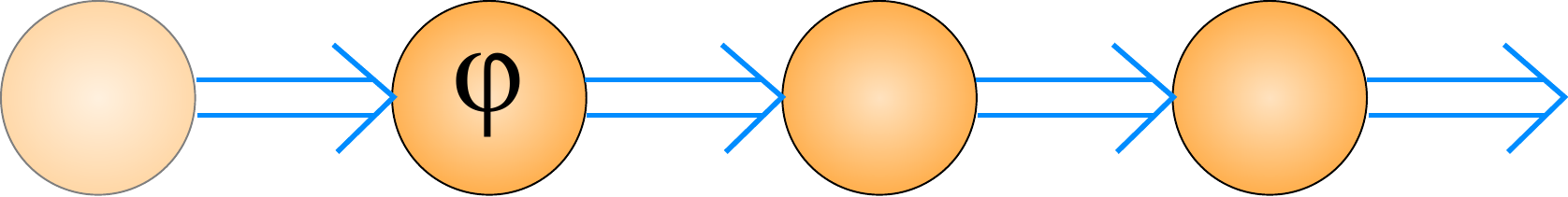}  
\end{center}
\end{minipage}
\begin{minipage}{0.52\columnwidth}
\begin{center}
\includegraphics[width=\columnwidth]{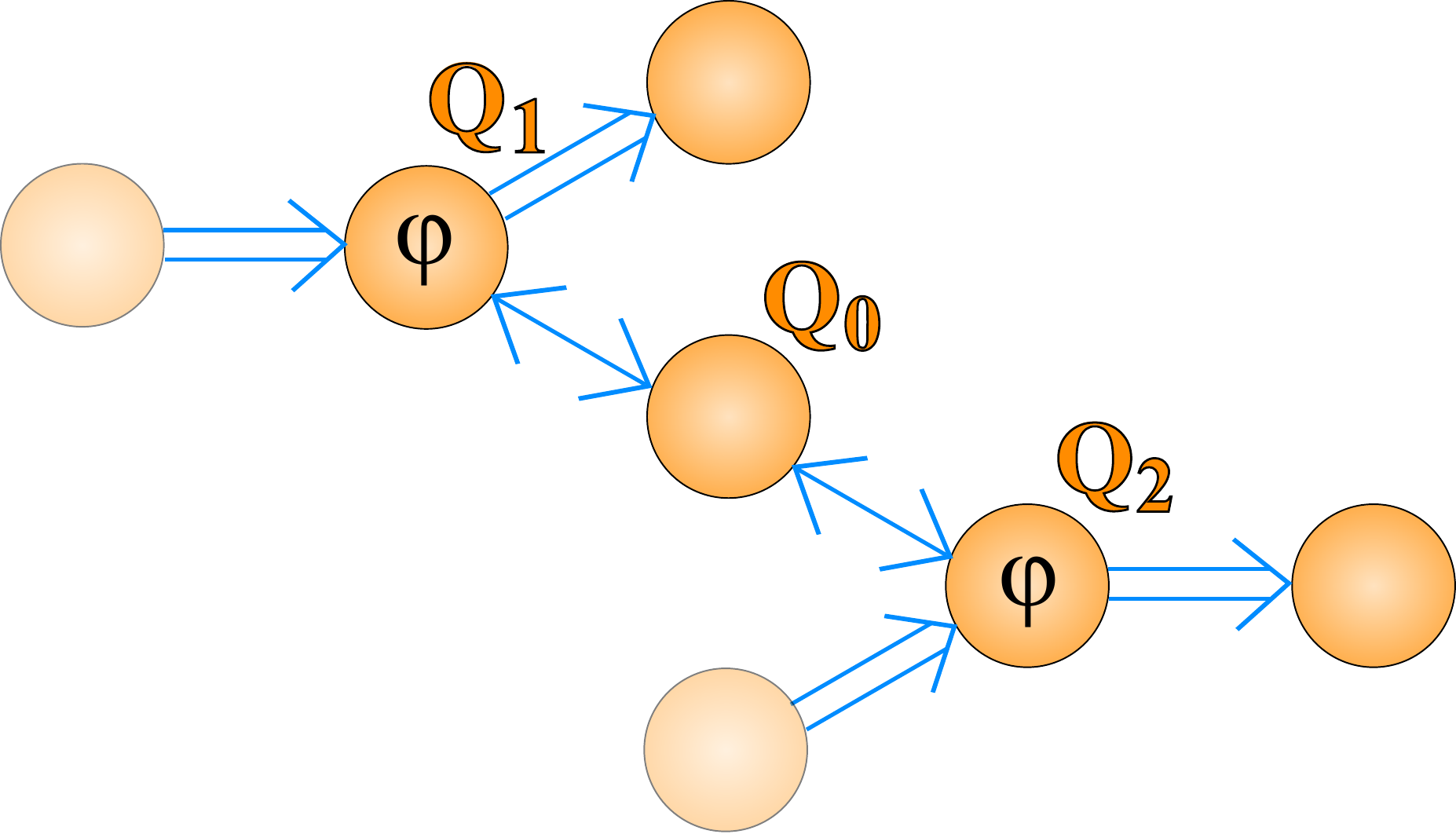}  
\end{center}
\end{minipage}
\caption{(Left) The timeline state and its graphical representation when $m = \tfrac{\pi}{4}$ or $\tfrac{3\pi}{4}$.
An intermediate state $\ket{\Psi^{\Rightarrow} (\varphi_\kappa)}$ is defined by having $\ket{\varphi_\kappa}_1$
at the current (leftmost) status instead of the fiducial state $\ket{o}_1$ in Eq.~(\ref{timeline}).
The double arrow $\Rightarrow$ represents the time direction realized by sending the 2 bits in the same direction,
(Right) The partially ordered structure which provides the spatial interactions, in addition to the timelines.
Given every bond which represents the projector $\Xi$ , we prescribe that the relatively left qubit is acted by the term $R^{z} (t)$ 
of Eq.~(\ref{Xi}) and the relatively right one is acted by $R^{w} (-t)$.   And we apply the projectors $\Xi$'s sequentially from the left
in a similar manner with the timeline state.
In other words, for each qubit, the bonds on the left-hand side are applied first, followed by the ones on the right-hand side. 
Note that the bonds in the same side commute each other and thus the ordering among them does not matter.
In contrast to the time direction by $\Rightarrow$,
the two-sided arrow $\leftrightarrow$ represents the space direction realized by sending the 2 bits in the opposite directions.
This way of communication is a ubiquitous feature of a measurement-based quantum computer, as particularly highlighted 
in the Figure 1 of \cite{miyake11}.
However, the quantum correlations here differ from those of the so-called 2D cluster state.
}
\label{fig:geometry}
\end{figure}

The strength of our approach gets further apparent when the {\it same} construction using our primitive correlation
is considered on a partially ordered set as the underlying geometry of the array of qubits.
The partial order may be constructed by the use of three projectors $\Xi$'s per qubit, extending the timeline state
by two projectors per qubit.
Note that this partial ordering is introduced to determine systematically the ordering of noncommutative $\Xi$'s, and 
it is per se different, for instance, from the causal set \cite{bombelli87} as a discretized Lorentian manifold.
We will demonstrate, in terms of the Figure~\ref{fig:geometry},  that the spatial interactions between independent degrees of freedom 
(namely the interaction terms in the conventional Hamiltonian formalism, in addition to the single-particle terms 
provided by the timeline state) can be realized as well.

For ease of the notation, 
let us denote an intermediate timeline state (i.e., the yet-to-be-unmeasured part) of $Q$ by 
$\ket{\Psi^{\Rightarrow} (\varphi_\kappa)}$, which is defined by
Eq.~(\ref{timeline}) with the replacement of the fiducial state $\ket{o}_1$ of the first unmeasured qubit by an arbitrary 
(not necessarily pure) state $\ket{\varphi_\kappa}_1$ representing the current status depending on $\kappa$.
The geometry by the right-hand side of Figure~\ref{fig:geometry} provides an additional action between two timelines, 
\beq
^{Q_0} \! \bra{\tau (\gamma)} \Xi_{s_{2}}^{\leftrightarrow} \Xi_{s_{1}}^{\leftrightarrow} 
\ket{\Psi^{\Rightarrow} (\varphi_{\kappa_1})}^{Q_1} \!\otimes \ket{o}^{Q_0} \!\otimes \ket{\Psi^{\Rightarrow} (\varphi_{\kappa_2})}^{Q_2} .
\eeq
(Note for a slight abuse of the notation that all $\Xi$'s act one after the other from the left of the Figure, so that indeed 
$\Xi^{\leftrightarrow}_{s_j}$'s act before the $\Xi$'s hidden inside $\ket{\Psi^{\Rightarrow} (\varphi_{\kappa_2})}^{Q_2}$.)
Integrating over $s_1$ and $s_2$ in a similar way with the analysis of the timeline state, 
we can show that the action is indeed a two-quibt unitary operation to the current statuses, namely
$U(\tau)  \varphi^{Q_1}_{\kappa_1} \otimes \varphi^{Q_2}_{\kappa_2}$,
where $U(\tau) = e^{-i \tfrac{\tau}{2}} \kb{0^z 0^w} + e^{i \tfrac{\tau}{2}} \kb{0^z 1^w } + e^{-i \tfrac{\tau}{2}} \kb{1^z 0^w }
- e^{i \tfrac{\tau}{2}} \kb{1^z 1^w}$.
It is always interacting, or entangling regardless of $\tau$, and particularly $U(\tau = 0, \pi)$ is the so-called Controlled-NOT gate.

The most remarkable feature of our model comes from how the distinction between temporal and spatial directions arises.
Scrutiny of the aforementioned interacting gate (c.f. Appendix~\ref{app:upsilon}) clarifies that one bit which marks a possible 
presence of $\sigma^{w}$ is sent down-rightward while another bit which marks that of $\sigma^{z}$ is sent up-leftward.
When the two bits are sent in the opposite ways, 
the relation of current statuses $\varphi^{Q_1}_{\kappa_1}$ and $\varphi^{Q_2}_{\kappa_2}$ are not causal (time-like), 
but effectively space-like via the common qubit $Q_0$.
Notably this distinction provides a marvelous solution to have the aforementioned area law of the conditional entropy
valid, even when the interactions between timelines are taken into account.
Namely, the amount of communication scales in proportional to the area size of the past region of qubits (multiple timelines), 
regardless of the time step, because the net amount of communication in the spatial directions is always zero.
This feature might be suggestive of the reason why time should be one dimensional.
While more spatial directions may be increased without the violation of the 1D area law, 
the communication amount cannot be constant per qubit in case the time directions are more than 1D.

Finally, let us briefly draw attention to that this ability to host the spacetime which respects our principle of least conditional entropy 
suffices to guarantee the computational universality as a quantum computer.
In particular, the unitary time evolution with an arbitrary $h^B$ in Eq.~(\ref{Sch}) is implementable in our operational 
model using digital quantum simulation.
Up to now, the pointers $\tau_{j}$ of the clock time (or the position-dependent potential term in the Dirac fermion picture) 
have been kept free. 
For example, it has been widely known that the set of elementary gates, composed by single-qubit rotations of arbitrary angles 
around two axes and a two-qubit gate such as CNOT is computationally universal.
Indeed, in addition to the CNOT constructed earlier, 
it is straightforward to show that such single-qubit rotations are realizable along 
every timeline using inhomogeneous $\tau_{j}$'s, while the least constant value of the conditional entropy is still intact.
The protocol would be analogous to a specific construction of measurement-based quantum computation using the 2D cluster 
state \cite{briegel01}, which has been analyzed in detail in \cite{raussendorf03} or recently in \cite{raussendorf11} 
with an analogy to the spacetime. 
We should note, however, that our quantum state on the partially ordered geometry is {\em not} a 2D cluster state which
is specified by the set of commuting stabilizer operators associated with the underlying graph.
Interestingly, in the 1D case the timeline state with $m = \tfrac{\pi}{4}$ or $\tfrac{3 \pi}{4}$ coincides with 
the 1D counterpart of the cluster state, known as the linear-graph state.
The Appendix~\ref{app:cluster} elaborates how our formulation using the time integrals can be related to the stabilizer formalism.  

At this point, it is worth noting a situation when $m$ is a multiple of $\tfrac{\pi}{2}$. 
Two axes commute as $\sigma^w = \pm \sigma^z$, and thus it is equivalent to the massless case ($m=0$). 
While this case attains $d^C = 2 = d^Q$, it can only provide a depolarization map on a {\em limited} set of the inputs.
Indeed, the quantum part $Q$ is always a GHZ-like state $\ket{0^z \cdots 0^z} + \ket{1^z \cdots 1^z}$ regardless of 
the underlying geometry. 
Curiously it has been shown in \cite{nest06} that this state does not provide computational complexity as powerful as
universal quantum computation.
So, one might be opt to say that if there were no missing information about the past (i.e., if the conditional entropy in 
Eq.~(\ref{principle}) were zero), the universe would rather miss all the richness of our world.

\section*{Summary}

We have explored a mechanism of time from a possible perspective that laws of nature are
constraints about how quantum and classical information can be processed.
In our formulation, a suitably correlated quantum system serves as a collection of internal clocks 
which defines not only time but also space via their relative phases.
Here emergent time is viewed as being made of continuous (quantum) phases which fill shades between 
discrete (classical) ordering related by classical communication.
The mechanism is driven by an innate balancing of entropies under sequential read-out of these clocks.
To regulate physically the amount of classical communication about past events in a steady manner, 
we have proposed the principle of least conditional entropy which states that the quantity is constrained to be positive.
It would be also intriguing to investigate how our approach could face general relativity,
particularly motivated by the entropic derivation of the Einstein equation \cite{jacobson95}. 
A wild speculation along this line is that the positivity of the conditional entropy, for instance, might be linked with 
that of the cosmological constant.


I acknowledge helpful discussions with W.K.~Wootters.
The work at Perimeter Institute is supported by the Government of Canada through Industry Canada and by Ontario-MRI.

\newpage
\appendix

\section*{Supplementary Information}

\section{Details of $\Upsilon^{CMQ}$}
\label{app:upsilon}

Conceptually it is only significant that $\Upsilon^{CMQ}$ is reversible locally, so that it only 
exchanges information among these degrees of freedom inside the site.
However, the details of processing $\Upsilon^{CMQ}$ at every single site for the simplest solution 
may be helpful to confirm how our mechanism of time works in a concrete fashion.
The protocol itself is analogous to that of the measurement-based quantum computer (MQC) using the 2D cluster state 
so that one could further consult its extensive references. (Note, however, that the formalism
as well as settings differ in several important details. In particular, classical information is usually handled 
{\em globally} in MQC, so that mixedness and thus entropies would not be addressed.)

The $\Upsilon^{CMQ}$ consists of two actions; a measurement of $Q$ and the storage of its outcome in $M$,
followed by a construction of the output information based on available classical information in $C$ and $M$.
So we denote $\Upsilon^{CMQ} = \upsilon^{CM}\upsilon^{MQ}$. 
First, for the timeline state with $m = \tfrac{\pi}{4}, \tfrac{3 \pi}{4}$ (i.e., $\sigma^w = \pm \sigma^x$), the signaling information 
is 2 bits, which are here labelled as $\kappa = (\kappa^z, \kappa^x)$ for the input $C$ and $\kappa' = ({\kappa^z}' , {\kappa^x}')$
for the output $C'$.
Then let us define
\begin{align}
\upsilon^{MQ} &=  {\bf 1}^{M} \otimes \kb{\tau(0)}^{Q} + \sigma^{x \, M} \otimes \kb{\tau(1)}^{Q} , \nonumber\\
\upsilon^{CM} 
& \left\{\begin{array}{ll}
{\kappa^z}' := \kappa^x  &\\
{\kappa^x}' := \kappa^z + \gamma  & \quad (\mbox{modulo 2}). \\
\gamma' := \gamma    &
\end{array} \right. 
\end{align}
Although it is not stressed in the main text for simplicity, the action of the measurement could depend not only on 
its outcome $\gamma$ but also on the input information $\kappa$. In case the clock time $\tau$ is not necessarily zero, we also set
$\tau (0) := (-1)^{\kappa^x} \tau$ and $\tau (1) := \tau(0) + \pi$ in $\upsilon^{MQ}$.
Also recall our construction that two measurement outcomes, $\tau (0)$ and $\tau (1)$, are equally likely, so that
$\gamma$ is always a totally random variable.
Suppose the input information $\kappa^z$ and $\kappa^x$ are totally random variables, then the output information
${\kappa^z}'$ and ${\kappa^x}'$ are totally random as well, according to $\upsilon^{CM}$.

Now it is straightforward to check that the above $\Upsilon^{CMQ}$ induces the following local processing, in correspondence
to the abstract expression of Eq.~(\ref{C'Q'}) given in the text.
Suppose initially $CQ$ is the form of Eq.~(\ref{CQ}) and its quantum part is $\ket{\Psi^{\Rightarrow} (\varphi_{\kappa})}^Q$. 
The dependence of $Q$ to the input information $C$ is given by 
$\varphi_{\kappa} = (\sigma^x)^{\kappa^x} (\sigma^z)^{\kappa^z} \varphi$,
where $\varphi$ is an arbitrary (not necessarily pure) single-qubit state which is independent of $\kappa$.
Then, for each $\kappa$ in Eq.~(\ref{CQ}),
\begin{align}
& \ket{\kappa}^C  \otimes \ket{0^z}^M \otimes  \ket{\Psi^{\Rightarrow} (\varphi_{\kappa})}^Q \nonumber\\
& \stackrel{\upsilon^{MQ}}{\longrightarrow} \ket{\kappa}^C \otimes \left(\ket{0^z}^M \otimes \ket{\tau (0)}\langle \tau (0) |\Psi^{\Rightarrow} (\varphi_{\kappa}) \rangle^Q  \right. \nonumber\\
& \left. \qquad + \ket{1^z}^M  \otimes \ket{\tau (1)}\langle \tau (1) | \Psi^{\Rightarrow} (\varphi_{\kappa})\rangle^Q \right) \nonumber\\
& \stackrel{\upsilon^{CM}}{\longrightarrow} \sum_{\kappa' \sim \gamma'} \ket{\kappa'}^{C'} \!\otimes 
\ket{\Psi^{\Rightarrow} (\varphi'_{\kappa'})}^{Q'} \!\otimes  \ket{{\gamma'}^z}^M \!\otimes \ket{\tau (\gamma')}^{\bar{Q'}} , 
\end{align}
where $\varphi'_{\kappa'} = (\sigma^x)^{{\kappa^x}'} (\sigma^z)^{{\kappa^z}'} A \varphi $.
In the second transformation $\upsilon^{CM}$, we have used the property of the transfer operator $A$
as defined in Eq.~(\ref{transfer}), and have changed the coordinate of classical information from $(\kappa, \gamma)$
to $(\kappa', \gamma')$.
Note that $A = VR^z (-\tau)$ now does not depend on any classical information, 
and the value of $\kappa'$, precisely ${\kappa^x}'$, is correlated with that of $\gamma'$ for fixed $\kappa$, as 
the notation $\kappa' \sim \gamma'$ implies.
In tracing out $M \bar{Q'}$ (i.e., the dependence on $\gamma'$) and taking all possibilities of $\kappa$ into account, 
we can reach a recursive form as presented in Eq.~(\ref{C'Q'}).

Next, when we consider the geometry by a partially ordered set, the role of $\Upsilon^{CMQ}$ per site stays qualitatively 
the same.
Three sites $Q_j$ ($j=0,1,2$) of the right-hand side of Figure~\ref{fig:geometry} should exchange 
classical information as follows, by modifying suitably their own $\upsilon^{CM}$'s.
\begin{table}[h]
{\tabcolsep = 0.4cm
\begin{tabular}{c|c|c}
\hline\hline
             & input  & output \\
\hline
$Q_0$  & $\begin{array}{ll} \kappa^z_0  & (\leftrightarrow) \\ \kappa^x_0 & (\leftrightarrow) \end{array} $     
             & $\begin{array}{ll} {\kappa^z_0}' := \kappa^z_2 & (\leftrightarrow) \\ 
             {\kappa^x_0}' := \kappa^x_1 + \gamma_0 & (\leftrightarrow) \end{array}$    \\
\hline
$Q_1$  & $\begin{array}{ll} \kappa^z_1 & (\Rightarrow) \\ \kappa^x_1 & (\Rightarrow) \\ {\kappa^z_0}'  & (\leftrightarrow) \end{array} $ 
 	    & $\begin{array}{ll}  {\kappa^z_1}' :=  \kappa^x_1 & (\Rightarrow) \\ 
	        {\kappa^x_1}' := \kappa^z_1 + {\kappa^z_0}' + \gamma_1 & (\Rightarrow) \\ 
	        \kappa^x_0 := \kappa^x_1 & (\leftrightarrow) \end{array} $ \\
\hline             
$Q_2$  & $\begin{array}{ll} \kappa^z_2 & (\Rightarrow) \\ \kappa^x_2 & (\Rightarrow) \\ {\kappa^x_0}'  & (\leftrightarrow) \end{array} $ 
 	    & $\begin{array}{ll} {\kappa^z_2}' :=  \kappa^x_2 + {\kappa^x_0}' & (\Rightarrow) \\ 
	        {\kappa^x_2}' := \kappa^z_2 + \gamma_2  & (\Rightarrow) \\ \kappa^z_0 := \kappa^z_2 & (\leftrightarrow) \end{array} $ \\
\hline\hline
\end{tabular}}
\end{table}

\noindent
As explicitly seen here, a new key change is the way of communication of classical information
while $d^C = 4$ is maintained among any neighboring pair of qubits.
Namely, as opposed to the 2-bit directional communication along a single timeline state, 
$Q_0$ receives one bit each from $Q_1$ and $Q_2$, and then sends out another bit each to them.
As discussed in the main text, this non-directional communication results in a realization of the spatial
direction in our model.

\section{1+1D Dirac fermion as clocks}
\label{app:dirac}

Our timeline state of Eq.~(\ref{timeline}) is partially motivated by an informal analogy to the 1+1D Dirac fermion.
Let us consider, in Eq.~(\ref{Sch}), the Dirac Hamiltonian of a relativistic fermion, which acts on the motional degree 
on the 1D spatial line parametrized by $r \in {\mathbb R}$ and the internal spin $\tfrac{1}{2}$ degree.
\beq
h^{B} = c(\hat{p} - \tfrac{\hat{A} (r)}{c}) \sigma^z + mc^2 \sigma^y ,
\label{dirac}
\eeq
where $\hat{p} =  -i \tfrac{\partial}{\partial r}$ is the momentum operator, $\hat{A}(r)$ is the vector potential depending on $r$,
$m$ is the rest mass, and $c$ is the speed of light which is set to be 1 hereafter.
It is widely known that the relativistic wave packet, initially localized well at the origin, spreads 
near the light cone at the speed of light (namely $\tfrac{c^2 \hat{p}}{h^{B}}$), so that the linear spreading on the position 
coordinate serves as keeping track of time (if it were able to be monitored).
It is crucial that two degrees of freedom are intrinsically entangled. 
Although the distribution of the spatial degree has been analyzed much in literatures, we are here interested in the evolution 
of the spin $\tfrac{1}{2}$ degree relative to that of the spatial degree.

Now the unitary evolution by Eq.~(\ref{dirac}) is given by the repetitive application of an elementary action
$e^{i \epsilon \hat{p} \sigma^z} e^{i \epsilon m  \sigma^y} e^{-i \epsilon \hat{A}  \sigma^z}$ at a discretized 
infinitesimal time unit $\epsilon$.
Indeed this can be see as a discrete-time quantum walk \cite{strauch06}, where its walker and quantum coin
correspond to the spatial and spin degrees of the Dirac fermion, respectively.
We realize that the sequential application of the transfer operator $V R^{z}(-\tau(\gamma))$ in Eq.~(\ref{transfer}) 
would lead to an analogous unitary evolution, by mapping operators of the Dirac fermion onto
the observable angles in the underlying Hilbert space of the qubit:
$ \hat{p} \leftrightarrow \theta_3, m \leftrightarrow \theta_2, \hat{A}(r) \leftrightarrow \tau(\gamma)$.
Note that originally the term by $\hat{p} \, \sigma^z$ is a {\em conditional} shift operator,  
moving to nearby orthogonal states of the spatial degree depending on the internal degree.
After mapping to a compact geometry of the qubit, however, there is only one orthogonal state among
the canonical clock-time states $\{\ket{t}\}$.
So two possible shifts, the anti-clockwise and clockwise rotations, become identical and actually have to move to 
the same orthogonal state by the $z$ rotation of $\theta_3 = \pi = -\pi$.
It is reminiscent of we can set $\theta_3$ to be $\pi (= -\pi)$ in the text, by different reasoning.
Observe also that $V$ is identical every step while $R^{z} (-\tau(\gamma))$ may depend on the site $j$,
which is the same way as the site-dependence of the Dirac Hamiltonian dynamics. 
So, not only has it motivated our timeline state, but also conversely it may suggest a renewed informational perspective 
such that the Dirac fermion describes correlations of internal clocks in a consistent manner with both quantum theory and
relativistic theory.

\section{Relation to the 1D cluster state}
\label{app:cluster}

We show that the timeline state of Eq.~(\ref{timeline}), which is the solution of Eq.~(\ref{principle}) 
with $m = \tfrac{\pi}{4}$ or $\tfrac{3 \pi}{4}$ (or $\sigma^w = \pm \sigma^x$),  satisfies simultaneously eigen equations,
\beq
K_j \ket{\Psi^{\Rightarrow} (\varphi)} = \ket{\Psi^{\Rightarrow} (\varphi)}  \; j = 2, \ldots , N ,
\eeq
regardless of $\varphi$.
Here we define $ K_j = \sigma^{z}_{j-1} \sigma^{x}_{j} \sigma^{z}_{j+1}$ for $j = 2, \ldots ,N-1 $ and 
$K_{N} = \sigma^{z}_{N-1} \sigma^{x}_{N}$.

The action of $\sigma^{z}_{j-1}$ on the $j-1$-th qubit results in $\sigma^{z} R^z (t_j) = (-i) R^z (t_j + \pi)$ in the integral of
Eq.~(\ref{timeline}). We introduce a new integration variable $\tilde{t}_j = t_j + \pi$, and see it affects to the $j$-th site 
located in the future of the timeline. By the action of $\sigma^{x}_j$ on this $j$-th site, we get
$\sigma^{x} R^z (t_{j+1}) R^x (-\tilde{t}_{j} + \pi) = i \sigma^{x} R^z (t_{j+1}) \sigma^{x} R^x (-\tilde{t}_{j}) = 
i R^z (-t_{j+1}) R^x (-\tilde{t}_{j})$, 
having the sign of $t_{j+1}$ now flipped.
Introducing another new variable $\tilde{t}_{j+1} = - t_{j+1}$, we see that, on the $j+1$-th site,
$\sigma^{z} R^z (t_{j+2}) R^x (\tilde{t}_{j+1}) \ket{o} = R^z (t_{j+2}) R^x (- \tilde{t}_{j+1}) \ket{o}$.
Thus, the action of $K_j$ brings the state of Eq.~(\ref{timeline}) back to the original one (using new 
variables $\tilde{t}_j, \tilde{t}_{j+1}$) with the eigenvalue $(-i) \times i = 1$.
That is how, the state corresponds to the so-called 1D cluster state \cite{briegel01} up to the freedom by $\varphi$,
in that a ``holographic'' degree of freedom on the left boundary is kept variable.
Indeed, for the fiducial case ($\varphi = o$) of Eq.~(\ref{timeline}) itself, 
$ \sigma^{z}_1 \ket{\Psi^{\Rightarrow} (o)} = \ket{\Psi^{\Rightarrow} (o)}$ is satisfied additionally, 
while the 1D cluster state should have another stabilizer by $K_1 = \sigma^{x}_{1} \sigma^{z}_{2}$.
Because of this correspondence, it is notable that this timeline state can be constructed 
without resort to its apparent ordering of $\Xi$'s,  by acting the commuting 2-qubit Controlled-Phase gate
$\kb{0^z 0^z} + \kb{0^z 1^z} + \kb{1^z 0^z} - \kb{1^z 1^z}$
parallelly on all the neighboring pairs of the qubits, following the initialization to a totally product state 
$\ket{0^z}_1 \ket{0^x}_2 \cdots \ket{0^x}_{N}$.
Needless to say, the same technique can be used to analyze the stabilizers for the case
of a partially ordered geometry.



\end{document}